\documentclass[epj,nopacs, referee]{svjour}
\usepackage{latexsym}
\usepackage{graphics}
\def\be{\begin{equation}}
\def\ee{\end{equation}}
\def\bea{\begin{eqnarray}}
\def\eea{\end{eqnarray}}
\newcommand{\il}{~}

\usepackage[centertags]{amsmath}
\usepackage{caption}
\usepackage{amssymb}
\newcommand{\ti}[1]{\mbox{\tiny{#1}}}
\newcommand{\api}[1]{\ ^{\mbox{\tiny{(#1)}}}\! }%

\begin{document}
%
%
\title{Stars in five dimensional Kaluza Klein gravity}
\author{D. Pugliese\inst{1,2} \and G. Montani\inst{3,1}
}                     
\offprints{}          
\institute{Dipartimento di Fisica, Universita' di  Rome, ``Sapienza'',  - Piazzale Aldo
Moro 5, 00185 Roma, Italy \and School of Mathematical Sciences, Queen Mary, University of London,
Mile End Road, London E1 4NS, United Kingdom.  \and ENEA-C. R. Frascati U.T.
Fus. (Fus. Mag. Lab.), via E.Fermi 45, I-00044, Frascati, Roma, Italy}
\date{Received: date / Revised version: date}
%
\abstract{
In the five dimensional Kaluza Klein  (KK) theory there is a well known class of  static and electromagnetic--free
KK--equations characterized  by  a naked singularity behavior, namely the Generalized Schwarzschild solution (GSS).
We present here a set of interior solutions  of five dimensional
KK--equations. These equations have been numerically integrated to  match the GSS in the vacuum.   The
solutions are candidates to describe the possible interior perfect fluid  source of the exterior
GSS metric and thus they can be  models for stars    for static, neutral astrophysical objects in  the ordinary (four dimensional) spacetime.
\PACS{
      {PACS-key}{discribing text of that key}   \and
      {PACS-key}{discribing text of that key}
     } 
} 
\maketitle
\section{Introduction}\label{sec:Intro}
Five dimensional (5D) original Kaluza Klein  (KK) model is one of the oldest multidimensional theory  providing a coupling between
gravity and electromagnetism~\cite{K1,K2,K3,Librogra}. The U(1) gauge invariance
arises as a spacetime symmetry. This is realized imposing the invariance under
translations on the compactified fifth dimension.
Multidimensional theories like String theory~\cite{JP}, Brane models~\cite{BM} and Supergravity~\cite{S}  arise for many respects  from the original KK--idea.
Great attention  is  devoted   to any  available tests
to probe the validity of these theories.
The study of the multidimensional  stellar structure could be  an interesting area of investigation of  the main astrophysical implications
of higher dimensional gravity. The analysis of the stellar structures,  arising from a model of gravity with one or more extra dimensions, might constitute therefore a natural arena to compare higher
dimensional theories of gravity with the  astrophysical phenomenology.
Generalized Schwarzschild solution  (GSS) is
a family   of free--electromagnetic, stationary 5D--vacuum
solutions of KK--equations.   In an asymptotic region of the class parameter the   four dimensional
counterpart metric of GSS is the Schwarzschild background.

The GSS  presents a naked singularity behavior that resolves in a black hole one only in the Schwarzschild limit~\cite{Overduin:1998pn,Wesson:1999nq,Lacquaniti:2009wc}.
But despite of the naked singularity nature showed far from the Schwarzschild
limit, GSS solutions are supposed to describe in principle the
exterior spacetime of any astrophysical sources that satisfy the
required metric symmetries.  Therefore many attempts have been made to study the proprieties of such spacetime ones  matter is included. In~ \cite{eqprinc}  an interior solution of the KK--equations  is considered to match, on the   effective exterior spacetime, the GSS.

In this paper we show a set of KK--interior solutions.
We consider in particular a free--electromagnetic, stationary, 5D--KK--model  within the compactification and cylindrical hypothesis. 4D--equations are  recovered after the KK--dimensional reduction of the 5D--equations  by the Papapetrou procedure.  Considering a perfect fluid 4D--energy--momentum tensor,  with  a polytropic  equation of state,  the solutions are recovered to match the GSS in the vacuum.
This paper is organized as follows: in Sec.\il(\ref{5DKKMODEL}) we will briefly introduce the KK--paradigm.
In Sec.\il(\ref{papaint}) we will face the revised approach to matter and dynamics in KK-theory. The GSS metric is introduced in Sec.\il(\ref{GSsSs}). The interior solution will be  discussed in Sec.\il(\ref{Interno}). Concluding remarks follow, Sec.\il(\ref{CCC3C})
\section{5D--KK--model}\label{5DKKMODEL}
We consider a 5D--compactified KK--model (see for instance \cite{Bailin:1987jd,Librogra}).
KK--spacetime is a 5D--manifold $\mathcal{M}^{\ti{5}}$,  direct product $\mathcal{M}^{\ti{4}}\otimes \mathcal{S}^{\ti{1}}$, between the ordinary 4D--spacetime $\mathcal{M}^{\ti{4}}$  and  the space--like loop  $\mathcal{S}^{\ti{1}}$.
In order to have an unobservable extra dimension  the  size on the fifth dimension is assumed to be
below the present observational bound or
\be\label{dalotano}
L_{\ti{5}}\equiv\int dx^{5} \sqrt{-g_{55}}<10^{-18}\rm{cm},
\ee
(Compactification hypothesis).
Condition (\ref{dalotano}) implies that  metrics components  do not depend
on the fifth coordinate (Cylindricity hypothesis): in fact it is  assumed  we are working  in  an
effective theory valid at the  lowest order of the
Fourier expansion along the fifth dimension \cite{Overduin:1998pn,Cianfrani:2009wj}.

Finally, it is  assumed that the $(55)$--metric component is a scalar.
Being $g_{55}=-\phi^2$ in Eq.\il(\ref{dalotano}),  the extra scalar field $\phi$  acts    as a  scale factor
governing the expansion of the extra dimension.
The KK--setup is characterized by the  breaking of the 5D--covariance and the 5D--equivalence principle \cite{Librogra}.
\section{Papapetrou's approach in KK--gravity}\label{papaint}
In~\cite{Lacquaniti:2009cr,Lacquaniti:2009rq,Lacquaniti:2009yy} the KK--field equations   in presence of a generic 5D--matter energy momentum tensor have been reformulated within the compactification hypotheses.
Extending  the cylindricity hypothesis to the 5D--matter tensor this becomes localized only in the ordinary 4D--spacetime but delocalized in the extra (compactified) dimension.

Then a Papapetrou multipole expansion \cite{papapetrou} of the  5D--energy momentum tensor is performed.
Performing   the dimensional reduction either for metric fields and
for matter ones  it is provided  a consistent approach  that removes the  problem of huge massive
modes.   Hence a consistent set of equations is wrote down for the complete dynamics of matter and fields;
with respect to the pure Einstein--Maxwell system there are now  two additional scalar fields: the usual KK--scalar field, $\phi$, plus a scalar source term, $\vartheta\equiv l_{\ti{(5)}} \phi T_{55}$, where $T_{55}$ is the $(55)$ component of the energy momentum tensor, given the coordinate length of the extra dimension $l_{\ti{(5)}}=\int dx^5$. In the cylindrical hypothesis it is $L_5=\phi l_{\ti{(5)}}$.
The simplest scenario, in which $\vartheta=0$, assures  the particles mass is constant and the
free falling universality of particles is recovered \cite{Lacquaniti:2009wc}.

In Sec.\il(\ref{sec:Papapetroumotion}) we will consider in details the Papapetrou approach to KK--particle dynamics, and in Sec.\il(\ref{sec:Papapetroumatter}) we shall briefly discuss the matter equations in the KK--scenario as developed by the revised approach \emph{a l$\acute{a}$}  Papapetrou.
\subsection{KK--particle dynamics}\label{sec:Papapetroumotion}
Matter dynamics in  KK--theories is generally faced
assuming, within the framework of the geodesic approach, the existence of a 5D--point--like
particle governed by a 5D--Klein--Gordon equation like\footnote{With Latin capital letters $A$ we
label the five dimensional indices, they run in $\{0,1,2,3,5\}$,
Greek  indices $\alpha$ run from 0 to 3,  $x^{5}$ is the angle
parameter for the fifth circular dimension.}
$P_{\ti{A}}P^{\ti{A}}=m_{\ti{5}}$, where $P_{\ti{A}}$ is the
5-momentum and $m_{\ti{5}}$ is  a constant mass parameter associated
to
the particle. Component $P_{\ti{5}}$ is conserved. 
Particle charge $q$ is defined
\be
q=\sqrt{4G}P_{\ti{5}}\quad\mbox{where}\quad
P_{\mu}P^{\mu}=m_{\ti{5}}^2+P_{\ti{5}}^2/\phi^2.
\ee
$G$ is the 4D--gravitational constant\footnote{We use units of $\hbar=c=1$.},  the  5D--velocities
$\omega^{\ti{A}}$ and 4D--velocities $u^{\ti{A}}$ are defined
respectively as
\be
\omega^{\ti{A}}\equiv \frac{dx^{\ti{A}}}{ds_{\ti{(5)}}},\quad
u^{\ti{A}}\equiv \frac{dx^{\ti{A}}}{ds},
\ee
with $\omega^{\ti{A}}=\alpha
u^{\ti{A}}$ and
\be
g_{\ti{A}\ti{B}}\omega^{\ti{A}}\omega^{\ti{B}}=1,\quad
g_{ab}u^{a}u^{b}=1,
\ee
where the $\alpha$ parameter reads $
\alpha\equiv ds/ds_{\ti{(5)}}$, and $ds$ ($ds_{\ti{(5)}})$ states for the 4D (5D)--line element.
The 5D--momentum reads
\be\api{5}P^{\ti{A}}\equiv
m_{\ti{(5)}}\omega^{\ti{A}}=m_{\ti{(5)}}\alpha u^{\ti{A}},
\ee
where
$
\api{4}P^{\ti{A}}\equiv m_{\ti{(5)}}u^{\ti{A}}.$
%
%

In \cite{Lacquaniti:2009cr,Lacquaniti:2009wc,Lacquaniti:2009rq}
is assumed that, as a consequence of the
compactification, test particles are described as a localized source
only in $\mathcal{M}^{\ti{4}}$ but are still delocalized along the fifth
dimension. KK--dynamics is therefore reformulated by a
multipole expansion \emph{a l$\acute{a}$} Papapetrou.

A 5D--energy--momentum tensor $^{\ti{(5)}}\!\mathcal{T}^{\ti{AB}}$ is associated
to a generic 5D--matter distribution governed by the conservation law $^{\ti{(5)}}\!\nabla_{\ti{A}} \
^{\ti{(5)}}\!\mathcal{T}^{\ti{AB}}=0$,  where $\partial_{5} \
^{\ti{(5)}}\!\mathcal{T}^{\ti{AB}}$=0.  Here $^{\ti{(5)}}\!\nabla$
($^{\ti{(4)}}\!\nabla$)is the covariant derivative compatible with
the 5D--metric (4D--counterpart metric $(ds)$).

Performing a multipole expansion of  $
^{\ti{(5)}}\!\mathcal{T}^{\ti{AB}}$ centrad on a four dimensional
trajectory $X^{\mu}$, such a procedure provides, at the lowest order, the
following equation:
\begin{equation}\label{padredimaria}
m u^{\mu}\,^{\ti{(4)}}\nabla_{\mu}u^{\nu}=
(u^{\nu}u^{\rho}-g^{\nu\rho})\left(\frac{\partial_{\rho}\phi}{\phi^{3}}\right)A+q
F^{\nu\rho}u_{\rho},
\end{equation}
$F^{\nu\rho}$ being the Faraday tensor, with the quantities
\begin{eqnarray}
m &=& \frac{1}{u^{0}}\int d^{3}x \sqrt{-g_4} \phi  \
^{\ti{(5)}}\!\mathcal{T}^{00},
\\\label{qdef}
q &=& e \hat{k} \int d^{3}x \sqrt{-g_4} \phi  \
^{\ti{(5)}}\!\mathcal{T}_{5}^{0},
\\\label{Adef}
A &=& u^{0} \int d^{3}x \sqrt{-g_4} \phi
^{\ti{(5)}}\!\mathcal{T}_{55},
\end{eqnarray}
%
where $\sqrt{-g_4}$ is the determinant of the 4D--metric,  $T^{AB}=\mathcal{T}^{AB}l_{\ti{(5)}}$ and $e\hat{k}$ reads $(e\hat{k})^2=\frac{4G}{c^2}$.
Eq.\il(\ref{padredimaria}) describes the   motion of a
point--like particle of mass $m$ in the 4D--spacetime, coupled to
electromagnetic field through charge $q$ and to the scalar field by
the new quantity $A$. The continuity equation
$\api{4}\nabla_{\mu}\api{4}J^{\mu}=0$, derived within the procedure
itself, implies that charge $q$ is still conserved.

It can be proved that in such a scheme the KK--tower of
massive modes is suppressed and that now particle's motion is governed by a
dispersion relation   $P_{\mu}P^{\mu}=m^2 $, where  dimensional
index $\api{4}$ is dropped. Nevertheless, as a consequence of the
coupling to the scalar field, the   mass $m$ is generally no more
constant, but is governed by the following equation\cite{Lacquaniti:2009wc,Lacquaniti:2009yy}:
\begin{equation} \frac{\partial m}{\partial
x^{\mu}}=-\frac{A}{\phi^{3}}\frac{\partial\phi}{\partial x^{\mu}}\, .
\label{padredisofia} \end{equation}
%
There exist a particular scenario  where
the conservation of the mass ($A=0$), or at least the validity of the free
falling universality ($A \infty m\phi^2$), is recovered. Topic is discussed in more details in
the cited works.

Particles characterized by $A=0$ just follows a geodesic equation:
\begin{equation}\label{PapA0}
u^{a}\,^{\ti{(4)}}\nabla_{a}u^{b}=0\, ,
\end{equation}
where even the usual scalar field coupling term disappears.
Therefore, these particles, whatever are their charges, turn out to be free also in $(\api{4}M,\api{4}g_{ab})$.
\subsection{KK--field equation}\label{sec:Papapetroumatter}
The full system of 5D--Einstein equation in presence of 5D--matter described by a   5D--matter tensor $\mathcal{T}^{AB}$ is:
\be
^5G^{AB}=8\pi G_5 \mathcal{T}^{AB},
\label{kkk}
\ee
where the following 5D--Bianchi identity holds
\be
^5\nabla_A T^{AB}=0,
\ee
while from the cylindricity hypothesis concerning the matter field it is
\be
\partial_5 T^{AB}=0,
\ee
where $G_5$ is the 5D--Newton constan,.
where $T^{AB}=\mathcal{T}^{AB}l_{\ti{(5)}}$, and  $G=G_5l_{\ti{(5)}}^{-1}$.
%
%
%
%
The components $T^{\mu\nu}$, $T_5^{\mu}$, $T_{55}$ are a 4D--tensor, a 4D--vector and a scalar respectively.
Introducing the quantities
\be
T^{\mu\nu}_{\ti{matter}}=l_{\ti{(5)}} \phi  \mathcal{T}^{\mu\nu}=\phi T^{\mu\nu}, \quad j^{\mu}=ek\phi T_5^{\mu},
\label{redef}
\ee
and $\vartheta=\phi T_{55}$, we have:
\be
\nabla_{\mu}j^{\mu}=0,
\ee
that introduces a conserved current $j_{\mu}$, related to the U(1) gauge symmetry and  coupled to the tensor $F_{\mu\nu}$, together with  the  conservation equation for  $T^{\mu \nu}_{\ti{matter}}$
\be
\nabla_{\rho}\left(T^{\mu\rho}_{\ti{matter}}\right)=-g^{\mu\nu}\left(\frac{\partial_{\rho}\phi}{\phi^{3}}\right)\vartheta+F^{\mu}_{\phantom\
\rho}j^{\rho},
\ee
coupled to the field $\phi$ and $A_{\mu}$ by the matter terms $T_{55}$ and $j_{\mu}$ and representing the energy--momentum density of the ordinary 4D--matter.

In the limit  $\phi=1$ we recover the conservation law for an electrodynamics system, and setting   $T_{55}=j^{\mu}=0$ we recover the conservation law $\nabla T^{\mu\nu}_{\ti{matter}}=0$.

%

Taking now into account these definitions, the reduction of Eq.\il(\ref{kkk}) leads to the set
\bea\label{ki}
G^{\mu\nu}&=&\frac{1}{\phi}
\left(\nabla^{\mu}\partial^{\nu}\phi+8\pi T^{\mu\nu}_{em} G -g^{\mu\nu}\Box\phi +8\pi G T^{\mu\nu}_{\ti{matter}}\right),\\
\label{kl} \Box\phi&=&\frac{8 \pi}{3}G
\left(T_{\ti{matter}}+2\frac{\vartheta}{\phi^2}\right)-G\phi^{3} F^{\mu\nu}F_{\mu\nu},
\eea
for the Einstein and KK--field,
where
\be\label{k12}
\frac{R}{2}=-\frac{3\phi^3(ek)^2F^{\mu\nu}F_{\mu\nu}}{8}
+8\pi G \frac{\vartheta}{\phi^3}.
\ee
The conservation equation becomes therefore
\be
\nabla_{r}\left(T^{\mu
\rho}_{\ti{matter}}\right)=-g^{\mu\nu}\left(\frac{\partial_{\rho}\phi}{\phi^{3}}\right)\vartheta+F^{\mu}_{\phantom\
\rho}j^{\rho},
\ee
where  the KK--Maxwell equation is now
\be
\nabla_{\nu}\left(\phi^3F^{\nu\mu}\right)=4\pi j^{\mu},
\ee
and
\begin{equation}\label{LAngelica}
T^{\mu\nu}_{\ti{matter}}\equiv \emph{l}_{\ti{(5)}} \phi
T^{\mu\nu},\quad j^{\mu}\equiv \emph{l}_{\ti{(5)}} \phi
T^{\mu}_{\phantom\ 5},\quad G =
\frac{G_{\ti{(5)}}}{\emph{l}_{\ti{(5)}}},\quad
\vartheta\equiv \emph{l}_{\ti{(5)}} \phi T_{55}.
\end{equation}
\section{Generalized Schwarzschild Solution}\label{GSsSs}
The GSS is a family   of free--electromagnetic, stationary 5D--vacuum
solutions of KK--equations.

Adopting  4D--spherical polar coordinates\footnote{Consider
$t\in\Re$, $r\in\left]2M,+\infty\right]\subset\Re^{+}$,
$\vartheta\in\left[0,\pi\right]$, $\varphi\in\left[0,2 \pi\right]$} $\left\{t,r,\theta,\varphi\right\}$ where
$d\Omega^2\equiv\sin^2\theta d\varphi^{2}+d\theta^{2}$ the GSS reads\footnote{Here $c=G=1$.}
\begin{equation}\label{CGMriunone}
ds_{\ti{(5)}}^{2}=\Delta^{\epsilon
k}dt^{2}-\Delta^{-\epsilon (k-1)}dr^{2}-r^{2}\Delta^{1-\epsilon
(k-1)}d\Omega^{2}-\Delta^{-\epsilon }(dx^{5})^2,
\end{equation}
where $\Delta=\left(1-2M/r\right)$, $M>0$ is a constant,
$\epsilon=\left(k^{2}-k+1\right)^{-1/2}$ and $k$ is a positive
dimensionless parameter. Each values of the $k$ sets a specific metric tensor and it characterizes the spacetime external to any astrophysical object.

On the spacetime section $dx^{5}=0$, the Schwarzschild solution is recovered for
$k\rightarrow\infty$, where in this limit $M$ is the Schwarzschild mass.

GSS metrics are asymptotically flat and,  for each finite value of $k$, GSS has a naked singularity situated in $r=2M$,
where the 5D--Kretschmann scalar and the square of the 4D--Ricci tensor diverge.

\begin{figure}
\resizebox{0.5\textwidth}{!}{%
  \includegraphics{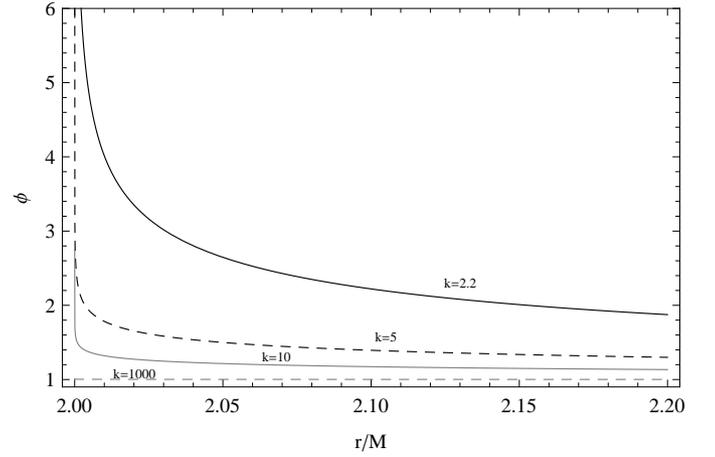}
}
\caption{The extra-dimensional scale factor $\phi=\sqrt{-g_{55}}$ as function of $r/M$, for different values of $k$.}
\label{Plotphi}       
\end{figure}

Moreover,  as $r$ approaches $2M$, $g_{tt}$ reduces to zero  while $g_{55}$ explodes to infinity, therefore
the  length of the extra dimension increases as well as $r$
approaches $2M$. It is possible to see that, at decreasing values of $k$, at fixed $r/M$ the 5D-scale factor  $\phi$ increases\footnote{However as $r\to 2\left(1+10^{-6}\right)$, for example, $\phi(k=2.2)=37.3618$.}, Fig.\il(\ref{Plotphi}).

%

Metric shows a pathological behavior as $r=2M$.
To investigate the nature of such a pathology we consider the values of one of the  curvature
invariants in that point. First we consider what happens in the five
dimensional spacetime. The
Kretschmann  scalar $\api{5}K_1\equiv R_{\ti{ABCD}}R^{\ti{ABCD}}$ is
$$\api{5} K_1=\frac{48 M^4 }{r^8}\left(1-\frac{2 M}{r}\right)^{2\epsilon(k-1)-4}$$$$
\left[\frac{r}{M}-1+\frac{\epsilon\left(\epsilon^2+1-2
k+\sqrt{\epsilon^4-2k\epsilon^2-1}\right)}{2} \right]
$$
$$
\left[\frac{r}{M}-1+\frac{\epsilon\left(\epsilon^2+1-2
k-\sqrt{\epsilon^4-2k\epsilon^2-1} \right)}{2}\right].
$$
This quantity diverges as $r\rightarrow2M$. This fact suggests
$r=2M$ as ``real'' singularity for the $\api{5}M$ endowed with metric
(\ref{CGMriunone}). The Kretschmann invariant $K_1$ of the Schwarzschild
spacetime is recovered
$$
K_1=\frac{48M^2}{r^6},
$$
as one takes the limit $k\rightarrow\infty$. Nevertheless to explore
the solution  behavior  for radius close to $r=2M$ in  the
corresponding  $\api{4}M$ , we calculated the Kretschmann scalar
$\api{4}K_1=R_{abcd}R^{abcd}$ relative to the ordinary four
dimensional spacetime
$$
\api{4}K_1=-\frac{24 M^4}{r^8} \left(1-\frac{2
M}{r}\right)^{2\epsilon (k-1)-4 } \left(\epsilon^2-2\right)$$
$$
\left(\frac{r}{M}-\frac{\epsilon\left[3+3\epsilon +2 k
\left(\epsilon
^2-3\right)\right]-6}{3
\left(\epsilon^2-2\right)}\right.$$$$\left.\frac{-\epsilon\sqrt{(2+k)\epsilon^4+(1-6k)\epsilon
^2-3}}{3
\left(\epsilon^2-2\right)}\right)
$$
$$
\left(\frac{r}{M}-\frac{\epsilon\left[3+3\epsilon +2 k
\left(\epsilon
^2-3\right)\right]-6}{3
\left(\epsilon^2-2\right)}\right.$$$$\left.\frac{+\epsilon\sqrt{(2+k)\epsilon^4+(1-6k)\epsilon
^2-3}}{3
\left(\epsilon^2-2\right)}\right).
$$
%
This quantity diverges for
$r\rightarrow2M$  for any fixed values of $k>0$. To evaluate the
balance between the two limits, one for large values of k--parameter
and the other  for r close to the Schwarzschild horizon  we also
made the series of $\api{4}K_1$ for $k\rightarrow\infty$ at some
fixed value of $r$, the first three  terms are
$$
\api{4}K_1=\frac{48 M^2}{r^6}-\frac{48 M^2 \ln\left[1-\frac{2
M}{r}\right]}{kr^6}-\frac{4 M^2(r-3M)\left(6r-11M\right)}{k^2r^6(r-2
M)^2}+$$
$$+\frac{12M^2 \ln\left[1-\frac{2 M}{r}\right] \left(2
\ln\left[1-\frac{2M}{r}\right]-5\right)}{k^2
r^6}+O\left[\frac{1}{k}\right]^3
$$
The Kretschmann
invariant $K_1$ of the Schwarzschild spacetime is recovered as one
takes the limit $k\rightarrow\infty$.
 The other
invariant we analyzed is the square of the Ricci tensor for the effective 4D--spacetime $\api{4}K_2\equiv R_{ab}R^{ab}$
$$
\api{4}K_2=-\left(1-\frac{2 M}{r}\right)^{2  \epsilon (k-1)}\frac{2 M^4 \epsilon^2}{r^4 (r-2 M)^4}
\left\{2 M r [3+\epsilon (2 k-3)]\right.$$\be\left. -3 r^2+M^2 \{\epsilon[6+2 k (\epsilon -2)-\epsilon ]-5  \}\right\}
\ee
this quantity reduces in
the Schwarzschild limit $\api{4}K_2=0$,  for  $k\rightarrow\infty$ at some
fixed value of $r$, the first  non vanishing term is
\be
\api{4}K_2=\frac{2 M^2 \left(9 M^2-10 M r+3 r^2\right)}{r^6 (-2 M+r)^2 k^2}+O\left[\frac{1}{k}\right]^3.
\ee
We have to conclude that the
point $r=2M$ is a physical singularity in the 5D--manifold as for
the ordinary 4D spacetime for all $k>0$.

The GSS naked singularity is surrounded by the induced scalar matter with a trace free energy momentum tensor.
The gravitational mass $M_g$ defined as \footnote{$dV_3$ is the ordinary
spatial 3D--volume element}
\be
M_g=\int(T^0_0-T^1_1-T^2_2-T^3_3)\sqrt{-g_4}dV_3
\ee
is, in the Schwarzschild limit, $M_g=M$; while, for each finite value of $k$, it  is $M_g=\epsilon
kM$ at infinity and  it goes to zero as $r$ closes the singularity.
\subsection{Particle motion}
The following conserved quantities characterize  test particle motion in a GSS spacetime:
$\mathcal{E}_{\ti{$\epsilon k$}}\equiv\xi^{a}_{(t)}P_{a}$,
$L_{\ti{$\epsilon k$}}\equiv\xi^{a}_{(\varphi)}P_{a}$, where
$\xi^{a}_{(t)}$ and $\xi^{a}_{(\varphi)}$ are  metric Killing
fields, or also
\begin{eqnarray}\label{Eq:qqq}
\mathcal{E}_{\epsilon k}\equiv\mathcal{E}\Delta^{k\epsilon-1},\;
L_{\epsilon k}\equiv L\Delta^{(1-k)\epsilon +1}\, ,
\end{eqnarray}
where $\mathcal{E}_{\ti{$\epsilon k$}}=\mathcal{E} $ and
$\mathcal{L}_{\ti{$\epsilon k$}}=\mathcal{L} $ in the
Schwarzschild's limit where
\begin{equation}\label{16}
\frac{E}{m}=\frac{(r-2M)}{\sqrt{r(r-3M)}};\quad
\frac{L}{m}=\sqrt{\frac{r^2 M}{(r-3M)}},
\end{equation}
we interpret $\mathcal{E}_{\ti{$\epsilon
k$}}$ and $\mathcal{L}_{\ti{$\epsilon k$}}$ as energy at infinity of
the particle and  total angular momentum, respectively. In terms
of these quantities   the  lagrangian density
$\mathcal{L}_{\ti{$\epsilon k$}}$ is defined as follows:
\begin{equation} \label{yyy}
\mathcal{L}_{\ti{$\epsilon k$}}\equiv -m^{2}\Delta^{-\epsilon (k-1)}
\left(\dot{r}\right)^{2}+\mathcal{E}_{\ti{$\epsilon
k$}}^{2}\Delta^{-\epsilon k} -L_{\ti{$\epsilon
k$}}^{2}r^{-2}\Delta^{\epsilon (k-1)-1}
\end{equation}
Here $\dot{r}\equiv u^{r}$, (see also\cite{Lacquaniti:2009wc} and \cite{Hackmann:2008tu,Enolski:2010if}.
%
\subsubsection{Equatorial orbits}\label{sssec:circular}
\begin{figure}
\resizebox{0.5\textwidth}{!}{%
\includegraphics{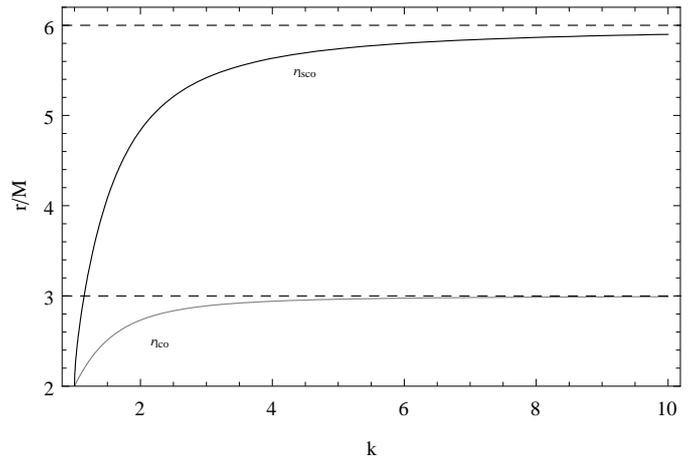}
}
\caption{Last circular orbits radius $r_{lco}/M$  (gray
line) and    last \emph{stable} circular orbit radius $r_{lsco}/M$
(black line) as functions of the $k$-parameter are plotted. For
large value of $k\geq1$ the last circular orbit radius approaches to
$r_{lco}=3M$ (its Schwarzschild's limit). Otherwise $r_{lco}<3M$:
circular orbits (unstable and stable) are possible also in a region
$r<3M$. Last stable circular orbit radius approaches $r_{lsco}=6M$
(its Schwarzschild's limit) for large value of $k\geq1$. Otherwise
$r_{lsco}<6M$: stable circular orbits are possible also in a region
$r<6M$.}
\label{RcEK}       
\end{figure}

The effective potential  $V_{\ti{$\epsilon k$}}\equiv E/m$, for massive
test particles in an equatorial orbit (with $k\geq1$) is defined as follows:
\begin{equation} \label{sss}
V_{\ti{$\epsilon k$}}\equiv\sqrt{ g_{\ti{00}} \left(1-
\frac{L_{\ti{$\epsilon k$}}^{2}}{m^{2}g_{\varphi\varphi}}\right)}\, .
\end{equation}
Studying $V_{\ti{$\epsilon k$}}$ as function of the orbit radius $r$,
we find the particle energy $\mathcal{E}_{\ti{$\epsilon k$}} $ and
the angular momentum $L_{\ti{$\epsilon k$}} $    of circular orbits: they are
%
\begin{eqnarray}\label{Dire}
\frac{\mathcal{E}_{\ti{$\epsilon k$}}
}{m}&=&\Delta^{\frac{k\epsilon}{2}}\sqrt{1+\frac{\epsilon
k}{\epsilon(1-2k)+\left(\frac{r}{M}-1\right)}},
\\\label{fare} \frac{L^{\pm}_{\ti{$\epsilon
k$}}}{mM}&=&\pm\sqrt{\frac{\epsilon k\frac{r^2}{M^2}
\Delta^{\left[(1-k)\epsilon+1\right]}}{\epsilon\left(1-2
k\right)+\left(\frac{r}{M}-1\right)}}.
\end{eqnarray}
\cite{Lacquaniti:2009wc}.
Last circular orbit radius  is
\begin{eqnarray}\label{Drr}
r_{lco}=\left[1+\epsilon(2k-1)\right]M\, ,
\end{eqnarray}
where  $r_{lco}<3M$ and approaches  this limit when $k$ goes to infinity
\cite{Lacquaniti:2009wc,Lacquaniti:2009yy}. On the other side,  as
$k$ approaches smaller values, last circular orbits radius approaches the
point $r=2M$ Fig.~(\ref{RcEK}); for $k=1$ it is $r_{lco}\equiv2M$.
This fact suggests to check for  circular orbits situated in
a region avoided by the 4D--Schwarzchild's physics. The aim of the
next part of this section is to explore the stability of such
orbits.
First, effective potential has two turning points
located in
$r^{\pm}/M=\left[1+\epsilon(3k-2)\pm\epsilon\sqrt{(k-1)(4k-1)}\right]M$
%
 where in the Schwarzschild's limit $r^{+}\equiv6M$ ad
$r^{-}\equiv2M$. We infer also from this fact that
last stable circular orbit radius is
\begin{equation}\label{xxx}
r_{lsco}=\left[1+\epsilon(3k-2)+\epsilon\sqrt{(k-1)(4 k-1)}\right]M\, .
\end{equation}
While in the Schwarzschild's limit we have $r_{lsco}  \equiv 6M$, here it is
$r_{lsco}<6M$ $\forall k>1$ and for $k=1$ we have $r_{lsco}=2M$; this
remarkably aspect of the circular motion in the GSS spacetime could
be
 a valid constraint to the theory   implying particles in
stable orbits for values of radius  just less than $6M$.
\begin{figure}
\resizebox{0.5\textwidth}{!}{%
\includegraphics{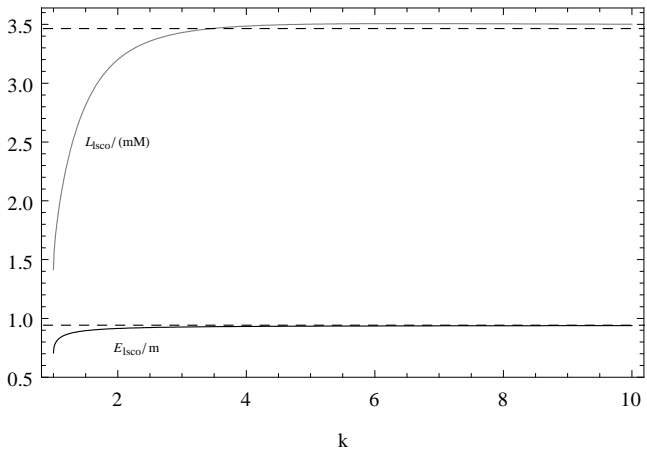}
}
\caption{The $\mathcal{E}_{lsco}/m$ (black line) and
$\mathcal{L}_{lsco}/m$ (gray line) in the circular orbits are
plotted as k-parameter functions. Schwarzschild' limits for the
energy and the angular momentum are also plotted (dashed lines). The
energy $\mathcal{E}_{lsco}$ is always under its Schwarzschild limit
while $\mathcal{L}_{lsco}$, for $k>3.45644$ is over the
Schwarzschild limit.}
\label{PLOTLEK}       
\end{figure}
%
Finally, another orbital deformation induced by the compactified
fifth dimension appears in the values of the energy and angular
momentum of the last stable circular orbits: the energy
$\mathcal{E}_{lsco}=\mathcal{E}_{\ti{$\epsilon k$}}(r_{lsco})$ is
always below its Schwarzschild's limit ($E_{lsco}=\frac{2
\sqrt{2}}{3}m=0.942809m$) for all values of $k\gtrsim 1$, while the
angular momentum $\mathcal{L}_{lsco}=\mathcal{L}_{ \epsilon k}(r_{lsco})$ is over the Schwarzschild's limit ($L_{lsco}=2\sqrt{3}Mm=3.4641Mm$) value for $k>3.45644$, Fig.~(\ref{PLOTLEK}).
\subsubsection{Radially falling   particles} \label{sssec:radial}
We consider here vertical  free falling test particles
in the GSS-background. To begin with, we note at first that the motion will be
described by $\dot{\theta}=\dot{\phi}=0$. Particle motion will be
regulated by the $t$-component and $r$-component of the geodesic equation
(\ref{PapA0}). However, radial motion has been extensively studied in
literature, see for example \cite{Overduin:1998pn,Kalligas:1994vf}.
We review here the main results and point out some considerations
based on the interpretation given by the Papapetrou's approach.
Particular attention is devoted to the dynamics in a region close to
the singularity ring $r=2M$.

\begin{figure*}
\resizebox{1\textwidth}{!}{%
\includegraphics{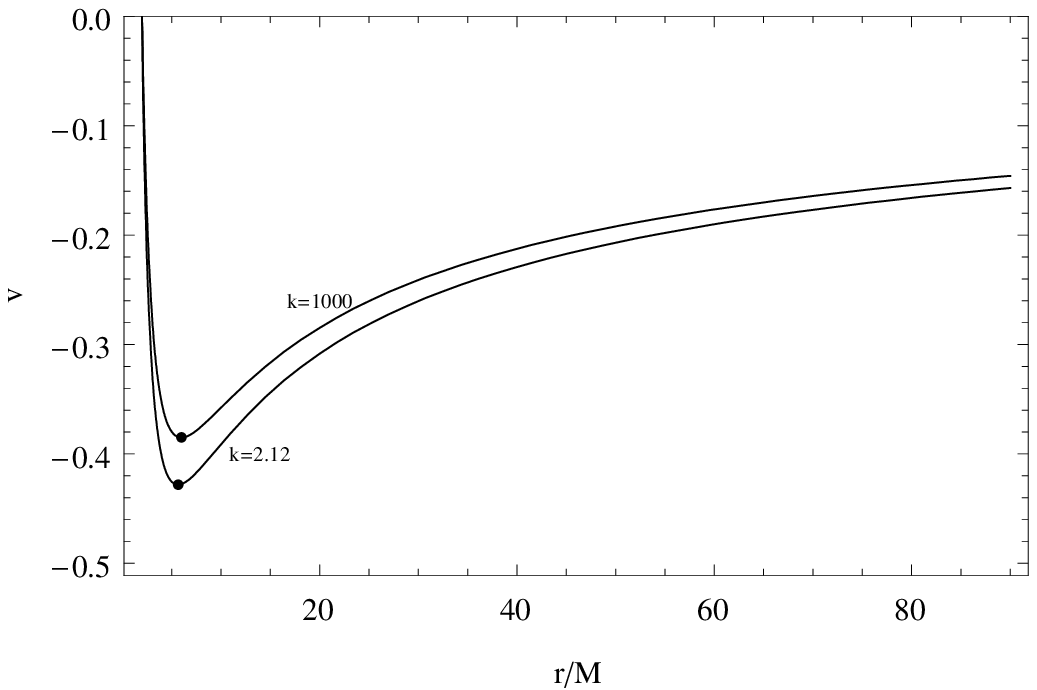}
\includegraphics{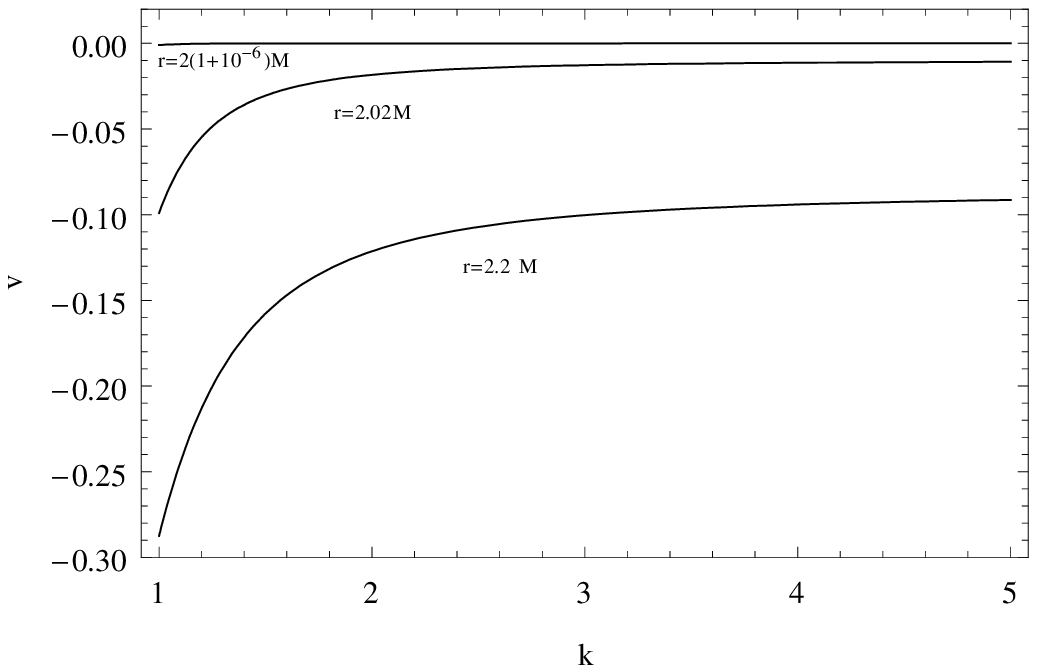}
}
\caption{The picture shows the coordinate velocity component in the
$r$-direction $v_{r}\equiv r'=dr/dt$, for a particle starting from a
point at (spatial) infinity as function of $r/M$ \emph{(a)} for
selected values of the k-parameter, and in \emph{(b)} as function of
the k-parameter at different values of $r/M$ close to $2M$.
}
\label{vkvak.eps}       
\end{figure*}

\begin{figure*}
\resizebox{1\textwidth}{!}{%
\includegraphics{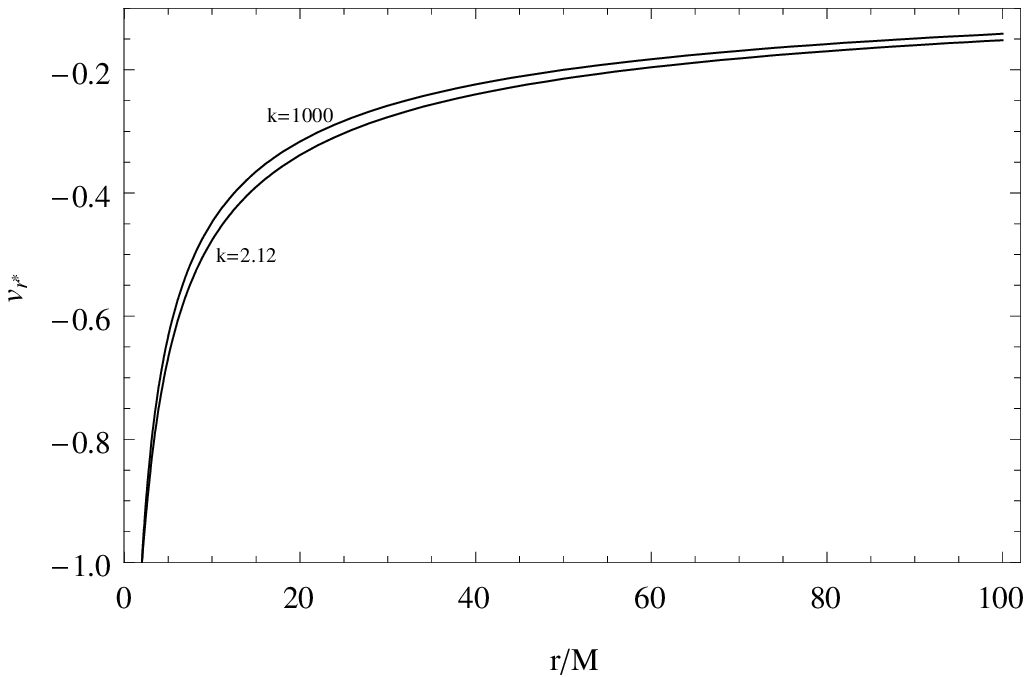}
\includegraphics{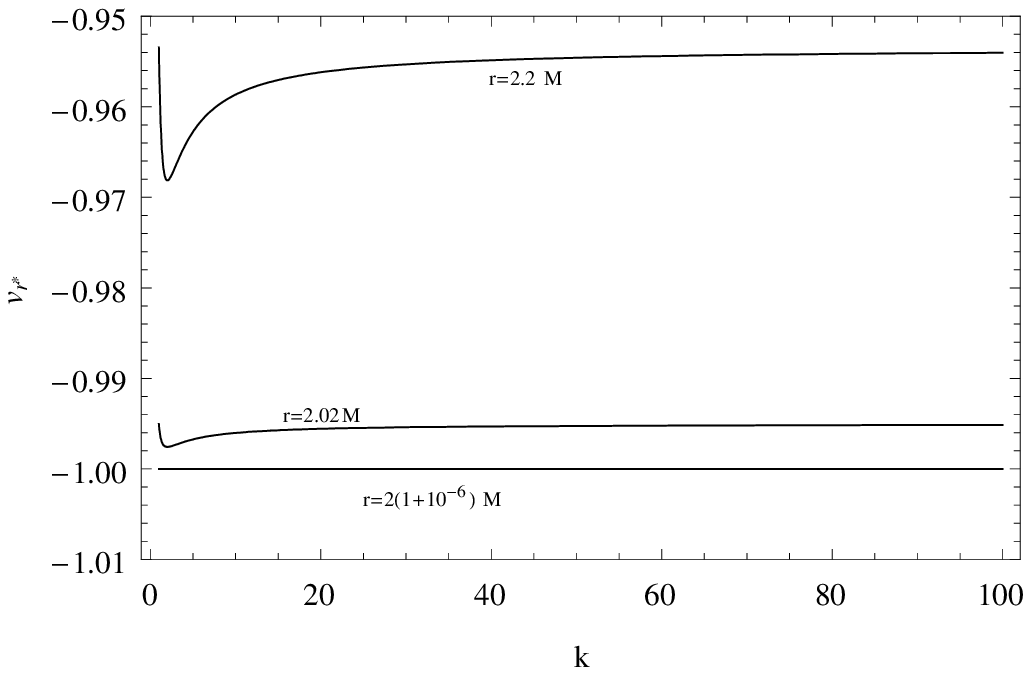}
 }  
\caption{The picture shows the locally measured radially velocity
$v_{r_*}\equiv r'_*=dr^*/dt$, where
$dr/dr^*=\sqrt{g_{tt}/g_{rr}}=\Delta^{\epsilon k-\epsilon/2}$, as
function of $r/M$ (left panel) for selected values of the k-parameter,
and in (right panel) as function of the k-parameter at different values
of $r/M$ close to $2M$.}
\label{vkvaklocal.eps}       
\end{figure*}

From (\ref{yyy}) we infer:
\begin{equation} \label{yyyradial}
\Delta^{-\epsilon k}\mathcal{E}_{\ti{$\epsilon k$}}^{2}
-m^{2}\Delta^{-\epsilon (k-1)} \left(\dot{r}\right)^{2}\equiv1\, .
\end{equation}
\cite{Overduin:1998pn}. The coordinate velocity component in the $r$-direction is
$v_{r}\equiv r'=dr/dt$. For a particle starting from a point at
(spatial) infinity we  have:
\be \frac{dr}{dt}=-\Delta^{\epsilon k-\epsilon/2}
\sqrt{1-\Delta^{\epsilon k}}\, . \ee
The locally measured radially velocity is $v_{r_*}\equiv
r'_*=dr^*/dt$ where $dr/dr^*=\sqrt{g_{tt}/g_{rr}}=\Delta^{\epsilon
k-\epsilon/2}$. Therefore we can write
\be \frac{dr^*}{dt}=-\sqrt{1-\Delta^{\epsilon k}}\, ,\ee
where $v_{r}$ and $v_{r^*}$ are function of $(r,k)$. Both converge
to zero at (spatial)-infinity. The radial velocity $v_{r}$ goes to
zero in the limit $r\rightarrow2M$, Fig.~(\ref{vkvak.eps}), and the local
measured velocity     $v_{r^*}$ goes to $(-1)$ in the limit
$r\rightarrow2M$, Fig.~(\ref{vkvaklocal.eps}).
%
%

\begin{figure}
\resizebox{0.5\textwidth}{!}{%
\includegraphics{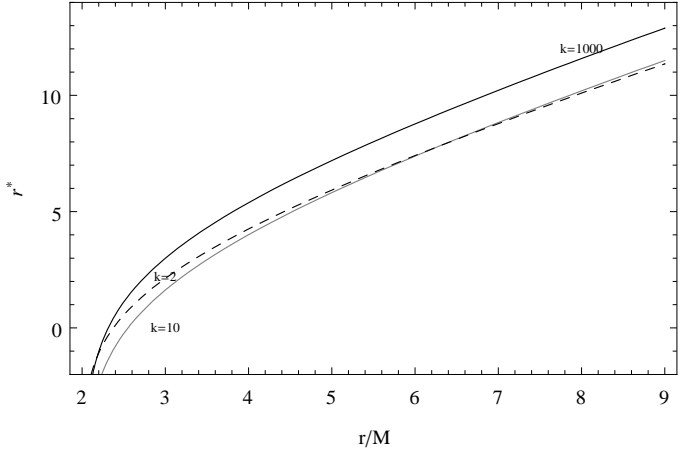}
}
\caption{The radius $r^*/M$ is plotted as function
of $r/M$ for selected values of $k$. Here
$dr/dr^*=\sqrt{g_{tt}/g_{rr}}=\Delta^{\epsilon k-\epsilon/2}$.}
\label{fig:pstar}       
\end{figure}

The evaluation for $r^*$ is given as follows:
\begin{eqnarray}\nonumber\frac{r^*}{M}=2\left(2-\frac{r}{M}\right)^{\frac{(2
k-1)\epsilon}{2}} \left(\frac{r}{M}-2\right)^{\frac{(1-2
k)\epsilon}{2}}\\ \label{rstars}
B\left[\frac{r}{2M},1-\frac{\epsilon}{2}+\epsilon
k,\frac{\epsilon(1-2 k+\frac{2}{\epsilon})}{2}\right]+C[k]\, .
\end{eqnarray}
%
We have the incomplete beta function $B[x,a,b]=B_x(a,b)$  where
$C[k]$ is a (imaginary)-constant of integration, to be fixed
in agreement to the Schwarzschild's limit and the reality  condition of the
solution. However,  $r^*$   always goes to infinity at
(spatial)--infinity. The picture is much more complex close to the
point $r=2M$; it strongly depends of the particular GSS solution,
 being $r^*(2M)$ a function of $k$. Indeed, for a fixed
value of $k$, $r^*(2M)$ is a negative number that approaches to
$(-\infty)$ as well as $k\rightarrow \infty$ Fig.\il(\ref{fig:pstar}).

%
%

\begin{figure}
\resizebox{0.5\textwidth}{!}{%
\includegraphics{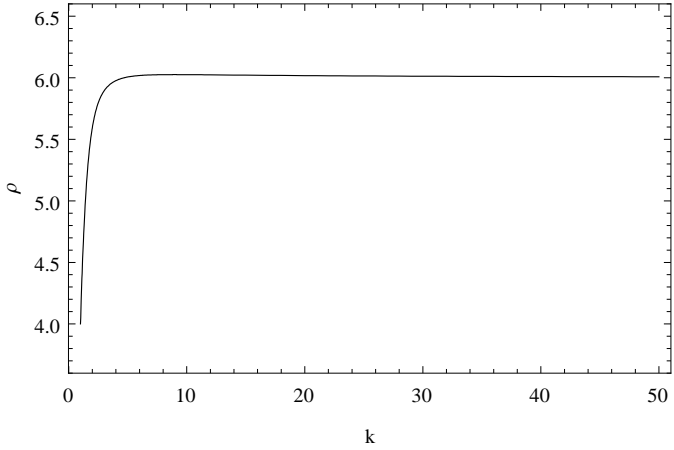}
}
\caption{The radius $\varrho$ at which the
radial velocity $r'$ starts to decreases is plotted as function of
k-parameter. In the Schwarzschild's limit it reduces to $\varrho=6M$,
otherwise $\rho=\rho(k)<6M$.}
\label{fig:Plotordunque}       
\end{figure}

%
In agreement  with \cite{Kalligas:1994vf} , the radius $\varrho$, at
which the radial velocity $r'$ starts to decreases, is
\be \label{varrho} \varrho=2M \left[1- \left(\frac{2k-1}{
3k-1}\right)^{1/(\epsilon k)}\right]^{-1}\, , \ee
that in the Schwarzschild's limit reduces to $\varrho=6M$, otherwise
we have $\rho=\rho(k)<6M$ Fig.\il(\ref{fig:Plotordunque}).
%
\section{The interior solution}\label{Interno}
The KK--equations (\ref{ki}--\ref{k12}) for  the  free electromagnetic case with $\vartheta=0$ are,
\begin{eqnarray}\nonumber
G^{\mu\nu}&=&\frac{1}{\phi}
\left(\nabla^{\mu}\partial^{\nu}\phi-g^{\mu\nu}\Box\phi +8\pi G T^{\mu\nu}_{\ti{matter}}\right),\\
\nonumber
&&\nabla_{\rho}\left(T^{\mu \rho}_{\ti{matter}}\right)=0,
\quad
R=0,\\
\label{klas8} \Box\phi&=&\frac{8 \pi}{3}
\left(T_{\ti{matter}}\right).
\end{eqnarray}
where $T_{\ti{matter}}\equiv g_{\mu\nu}T^{\mu \nu}_{\ti{matter}}$. GSS is a solution of eqs.~(\ref{klas8}) with $T^{\mu\nu}_{\ti{matter}}=0$.
We consider a  (4D) perfect fluid energy momentum tensor, $T^{\mu \nu}_{\ti{matter}}\equiv T^{\mu\nu}$:
\begin{equation}\label{Timpo}
T_{\mu\nu}=(p+\rho) u_{\mu} u_{\nu} -pg_{\mu\nu}.
\end{equation}
where $p$ is the pressure, $\rho$ the density.
Therefore eqs.~(\ref{klas8}) became
\begin{eqnarray}\label{kiastra1}
G_{\mu\nu}&=&\frac{1}{\phi}
\left(\nabla_{\mu}\partial_{\nu}\phi-g_{\mu\nu}\frac{8
\pi}{3}
\left(\rho-3p\right) +8\pi  T_{\mu\nu}\right),\\
\label{klastra2} \Box\phi&=&\frac{8 \pi}{3}
\left(\rho-3p\right), \quad
\nabla_{\rho}\left(T^{\mu \rho}\right)=0.
\end{eqnarray}

\begin{figure*}
\resizebox{1\textwidth}{!}{%
\includegraphics{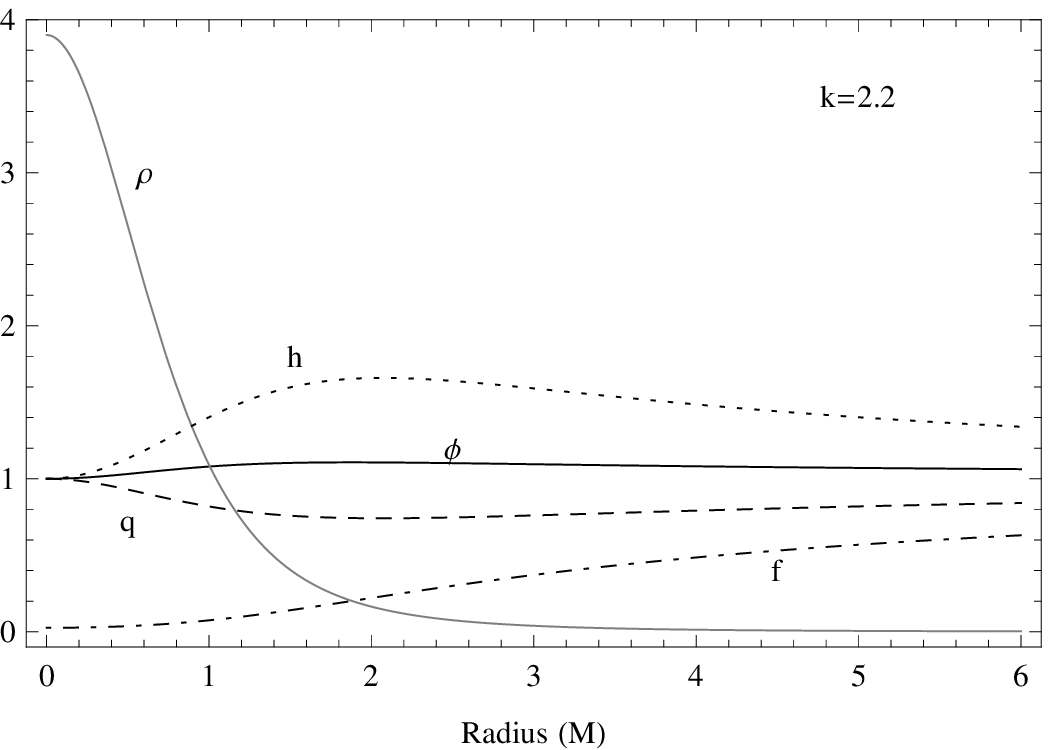}
\includegraphics{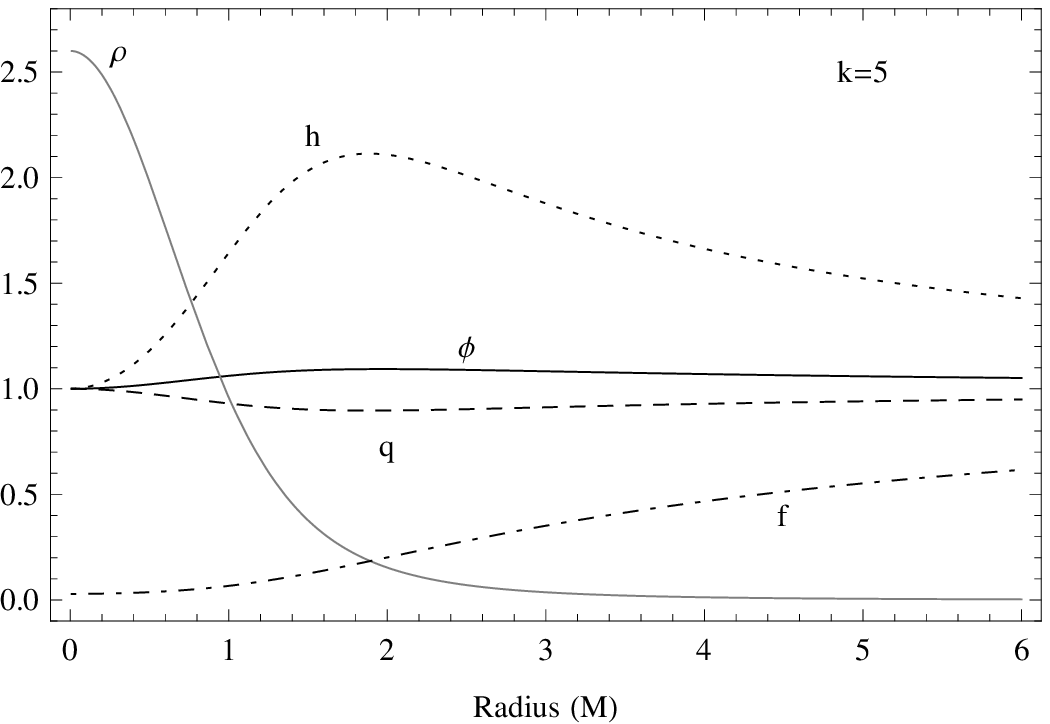}
\includegraphics{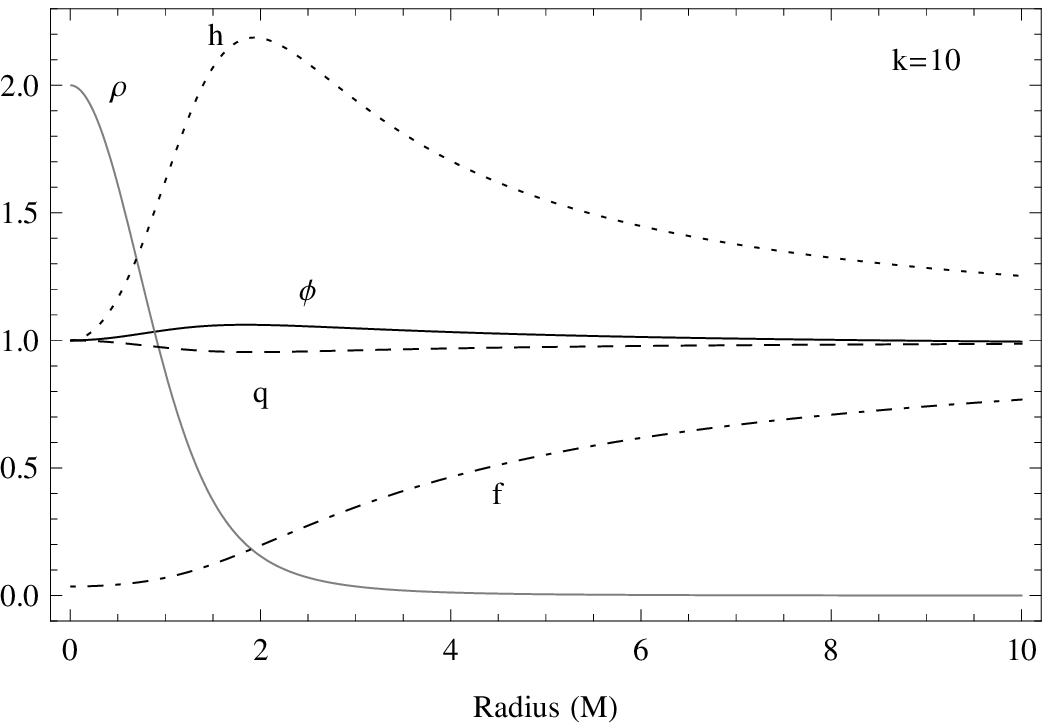}
\includegraphics{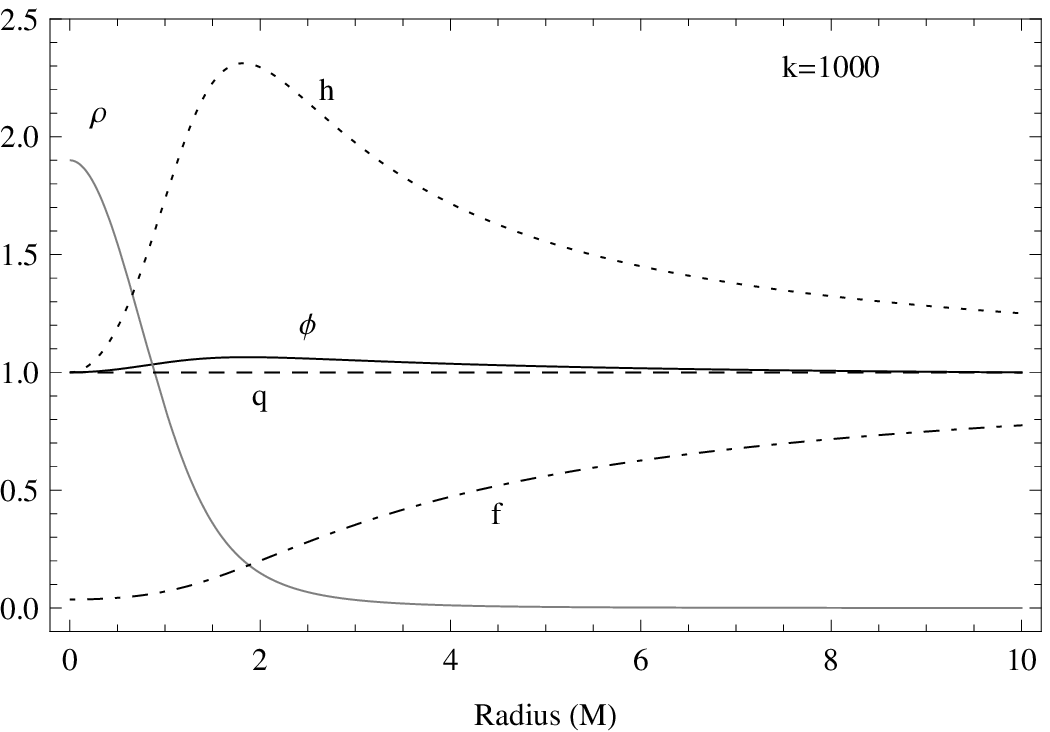}
 }    
\caption{Metric coefficients, $f$ (dotdashed line), $h$ (dotted line), $q$ (dashed line),  $\phi$ (black line) and the density $\rho$ (gray line), where $p=a\rho^b$, as functions of the radial coordinate, in unit of mass $M$, for different values of $k$, for $b=4/3$.
For $k=2.2$ (upper left panel), $x=1.7\cdot10^5$, $a=0.36$, $\rho_0=3.9$.
For $k=5$ (upper right panel), $x=9 \cdot 10^5$, $a=0.399$, $\rho_0=2.6$.
For $k=10$ (bottom left panel), $x=2\cdot10^4$, $a=0.4$, $\rho_0=2$.
For $k=10^3$ (bottom right panel), $x=300610$, $a=0.39$, $\rho_0=1.9$.}
\label{Fig:stellaRoma1}       
\end{figure*}

We look for a  solution of
eqs.~(\ref{kiastra1}, \ref{klastra2}) that matches the GSS on  a certain radius $r=R$ where  $p(R)=0$.
In order to do this we considered an  equation of
state $p=a\rho^b$ with $(a,b)$ constants and a
metric  solution of the form\footnote{Consider
$t\in\Re$, $r\in\left]0,+\infty\right]\subset\Re^{+}$,
$\vartheta\in\left[0,\pi\right]$, $\varphi\in\left[0,2 \pi\right]$}
\begin{equation}\label{MetriIntKK}
ds_{\ti{(5)}}^{2}=f dt^{2}-h dr^{2}-q d\Omega^{2}-\phi^2 (dx^{5})^2,
\end{equation}
where  $(f, h, q, \phi)$ are functions of  $r$ only.
From the conservation equation in (\ref{klastra2}) we obtain
\begin{equation}\label{fsol}
f(r)=c\left(\frac{\rho (r)}{\rho (r)+a \rho (r)^b} \right)^{\frac{2 b}{b-1}},
\end{equation}
where the constant  $c$ must be fixed to  match  the exterior solution(GSS).
Noting that the equations do not change under the transformation $f(r)\rightarrow f(r)/c$,   the solution (\ref{fsol})  is  substituted in the set
eqs.~(\ref{kiastra1}, \ref{klastra2}).
We integrated numerically   the three independent equations, the $(tt)$  $(rr)$ components of  the Einstein equations  and the Klein Gordon equation, for the variables $\Delta, \phi$, and $\rho$ where
\begin{equation}
h(r)=\Delta (r)^{-\epsilon (k-1)},\quad q(r)=\Delta (r)^{1-\epsilon k-1)},
\end{equation}
and $\Delta (r)$ is now an unknown function of the radial coordinate.
To  establish the initial conditions we  imposed
\begin{eqnarray}
\phi_0&\equiv&\phi(0)=\left(1-\frac{2}{x}\right)^{-\epsilon /2},\\
\phi'_0&\equiv&\phi'(0)=-\frac{\epsilon\left(1-\frac{2}{x}\right)^{-(1+\frac{\epsilon }{2})}}{x^2}.
\end{eqnarray}
%
Therefore, for fixed values of $a, b$ and $\rho_0$, we solved an ``eigenvalue" problem for $x$, where $x$ is  fixed  to verify the match conditions
\begin{eqnarray}
\Delta(R)&=&\left(1-\frac{2}{R}\right),\quad\Delta'(R)=\frac{2}{R^2},\\
\phi(R)&=&\left(1-\frac{2}{R}\right)^{-\epsilon /2},\\
\phi'(R)&=&-\frac{\left(1-\frac{2}{R}\right)^{-1-\frac{\epsilon}{2}}\epsilon}{R^2}.
\end{eqnarray}
Some solutions, for different values of $k$, are plotted in Fig.~(\ref{Fig:stellaRoma1}).

The index $b$ has been fixed $b=4/3$. Four different cases of $k\in\{2.2, 5, 10, 1000\}$ have been considered.
Thus, the Einstein-Klein-Gordon equations have been integrated for a fixed central  density $\rho(0)$.
In the recovered solutions the interior ordinary 4D matter is characterized by a polytropic equation of state $p(\rho)=a \rho^{4/3}$.    The integration procedure leads to a  constant $a\approx0.4$, for each values of $k$ and $\rho_0$.  The central density $\rho_0$ decreases increasing the $k$ parameter. In this configurations the scalar field  matter, solution of  massless  Klein-Gordon equation is coupled to this matter by Eq.\il(\ref{klastra2}).

The physical proprieties of each solutions  and the astrophysical implications of their  existence  requires a more detailed investigation and it will be discussed  elsewhere~\cite{next}.
\section{Furthers applications}
We conclude this discussion  briefly introducing the possible concrete applications of the analysis of the interior, stellar solutions associated to the GSS family of metrics.

Stellar models in extended or higher dimensional theories of gravity have been extensively studied in literature, in many contexts (see for example \cite{Capozziello:2011nr,FeKK,PoncedeLeon:2007mi,deLeon:2007qe,Patel:2001jw,cha,PaulBC,tot,tat}).

For example, it could be interesting to compare with other alternative proposals where stellar
solutions are derived. In view of
this,  we consider,  for instance, the results reported in \cite{Capozziello:2011nr}  in which  the hydrostatic equilibrium and stellar structure in $f(R)$ gravity has been studied. As well enlighted in \cite{Capozziello:2011nr} the strong gravity regime is a valuable  way to check the validity  of the extended and the multidimensional  theories of gravity, in
particular  the formation and the evolution of stars can be
considered suitable test for these theories.

The  study of stars in five dimensional Kaluza Klein gravity could develop in two main directions that are also  future developments on this work: the extensive study of the phenomena around the 4D stellar objects and the comprehension of the dynamics of such kind of objects.

In the compactification hypothesis is predominant the idea to find a possible field of phenomena in which  the
presence  of the  extra dimension
takes evidence. This fact has to be taken seriously into account especially  in those contexts in which the strong gravity effects could matches the physics of the Planck scale \cite{Capozziello:2010jx}. On one side,  attention is for example  devoted to the microphysics of the LHC to search for a sign of the Planck scale length dimension \cite{Kong:2010mh,Bhattacherjee:2010vm,Datta:2010us,Franceschini:2011wr}, but many efforts are also given to the comprehension of the higher energy astrophysical phenomena linked to the life steps of stellar evolution and that could be a natural arena in which explore the effects of a multidimensional theory \cite{KKcompa1,KKcompa2}, especially for those cases in which the standard models of stellar structure and evolutions does not properly collide with observed data; suggesting therefore a different or a modification of the basic mechanics that rules  stars matter and life \cite{Hannestad:2003yd}). An example can be provided by the  magnetars or anomalous neutron stars, (see for example \cite{Capozziello:2011nr} and reference therein).



The exploration of the motions in the vacuum spacetime, partially faced  here and in \cite{Lacquaniti:2009wc,Lacquaniti:2009rh} within the Papapetrou approach can be completed by  the  analysis  of the emission processes induced by charged particles in the GSS background. The comparison in  particular of the electromagnetic emission spectra  due to a freely falling charged test particle, and in general the test particle dynamics,   in the GSS and in Schwarzschild background could get light on physics around the singular ring $r=2M$, whose naked singularity nature is actually ambiguous (see also \cite{Virbhadra:2007kw,Virbhadra:2002ju} and \cite{lw,ww}).  However the solution of the equations governing the emission process should strongly depend on the boundary conditions imposed on the equations. A comparison between the electromagnetic emission by radially falling charged particles in  the GSS and stellar case could be strongly discriminant.

\section{Conclusions}\label{CCC3C}
The aim of this work was to verify  the existence of an interior solution  of the KK--equations to mach with  the vacuum spacetime of  the Generalized Schwarzschild solution characterized  by a naked singularity.  For this purpose  we solved
the set of five dimensional static,  electromagnetic-free KK--equations. These have been  numerically integrated  for a  (4D) perfect fluid energy momentum tensor with a polytropic equation of state.
These solutions are matched with  the Generalized Schwarzschild solution.

There is a great
interest in   providing a theoretical model able to explain the role
of the extra dimensions and their compatibility in a world that looks like a
four dimensional one.

Therefore  any experimental observation
compatible with such theories could be a strong
constraint concerning their validity for example   by exploring  the dynamical effects of the extra compactified
dimension.
Using an effective potential approach to the motion, we
showed the last circular orbit radius and in particular
the last stable circular orbits radius of a charged or neutral test
particles \cite{Lacquaniti:2009wc}.
The detailed study of the physical features on such astrophysical objects will be a future work.

The study of bounded  configurations in KK--paradigm is intriguing for many respects: from one side it could represent an environment in which an hypothetical extra dimension shows a recognizable  fingerprint on the dynamics and global properties of these objects. On the other side the interior solutions for the GSS--like spacetime  are particular interesting if the GSS is read as KK--generalization of a Schwarzschild metric for an electrically neutral spherically symmetric 4D--object. In fact,  if the final state of the  star evolution in the Schwarzschild spacetime could lead to a 4D--black hole, the final state of an neutral spherically symmetric  4D--object in GSS should lead to a naked singularity for the ordinary 4D--spacetime.   Nevertheless both the solutions describe neutral spherically symmetric  4D--object the difference being in the presence,  for the KK--solution, of a  coupling with a scalar matter.
In this respect, the comprehension of the mechanisms under these different evolutions of the
neutral spherically symmetric  4D--object, one in the general relativistic framework, and the  other with the contribution of a scalar field in the GSS, is challenging,
\cite{Yamada:2011br,deLeon:2010rp,Eingorn:2011vu,Eingorn:2010wi}.
%
%
\subsubsection*{Acknowledgments}
This work has been developed in the framework of the CGW Collaboration (www.cgwcollaboration.it). We would like to thank V. Lacquaniti for helpful comments on this topic.
One of us (DP) gratefully acknowledges financial support from the A. Della Riccia Foundation.
%
%
%
%

\end{document}